\documentstyle[subeqn]{article}

\begin{document}

\begin{flushright}
TITCMT-95-30
\end{flushright}
\begin{center}
{\Large\bf
Exact Ground-State Energy\\
of the Ising Spin Glass on Strips\\
}
\vspace{3ex}
{\large
Tadashi {\sc Kadowaki},
Yoshihiko {\sc Nonomura}\hspace{-3ex}\footnote{
Present address: Department of physics, University of Tokyo, Hongo,
Bunkyo-ku, Tokyo 113.
}
and Hidetoshi {\sc Nishimori}\\
}
\vspace{1ex}
\normalsize
Department of Physics, Tokyo Institute of Technology,\\
Oh-okayama, Meguro-ku, Tokyo 152
\end{center}

\abstract
{
We propose a new method for exact analytical calculation of the ground-state
energy of the Ising spin glass on strips.
An outstanding advantage of this method over the numerical transfer matrix
technique is that the energy is obtained for complex values of the probability
describing quenched randomness.
We study the $\pm J$ and the site-random models
using this method for strips of various sizes up to $5\times\infty$.
The ground-state energy of these models is found to have singular points
in the complex-probability plane, reminiscent of Lee-Yang zeros
in the complex-field plane for the Ising ferromagnet.
The $\pm J$ Ising model has a series of singularities which may
approach a limiting point around $p \sim 0.9$ on the real axis
in the limit of infinite width.
}

\section{Introduction}
{}~

In spite of long history of intensive studies, there are few exact
results known on models of spin glasses.
To summarize analytical calculations of the ground-state properties of
the Ising model,
Derrida {\it et al.}~\cite{Derrida} investigated linear chains,
double-chain strips and the $3\times3$ square lattice systems
using a variant of the transfer-matrix method to obtain exact expressions
of the ground-state energy.
Morita and Horiguchi~\cite{Morita} and
Fechner and B{\l}aszyk~\cite{Fechner}
investigated the linear triangular chain by summing up all
the configurations of frustrated clusters.
They derived analytical expressions of the ground-state energy for
these systems.
However, their methods are essentially an enumeration
of possible configurations
and therefore are very difficult to be generalized to systems with
larger widths.

Recently, interest in this problem has been renewed
by Dress {\it et al.}~\cite{Dress} who
applied the method of Derrida {\it et al.}~\cite{Derrida}
to the problem of error-correcting codes.
In the present paper, we generalize their method to the calculation
of the ground-state energy of the $\pm J$ model
and the site-random model on strips of various width.
An important reason to consider the latter site-random model is that
almost all experiments of spin glasses are
carried out on site-random systems.
Although bond-random and site-random models would share
many important properties,
there should exist some distinctions, a part of which is clarified
in the present paper.
An advantage of this method over the enumeration methods is that
ground-state energy of strips of various width
with various kinds of randomness
can be calculated systematically without elaborate enumeration
of configurations.

A useful aspect of the method used in the present paper is that
the ground-state energy is obtained for general complex values of parameters
to control probabilities of quenched random variables.
This is in sharp contrast to the numerical transfer matrix technique which
is applicable only to real probabilities.
As long as the width of a strip is finite, the ground-state energy
of models on the strip is analytic as a function of the probability of
quenched randomness if this probability is restricted to real values.
However, when the variable to control probability
is analytically continued into
the complex plane,
the ground-state energy is found to have singularities.
This analytic continuation into the complex plane is analogous to
the idea of Lee-Yang zeros~\cite{Lee-Yang} for complex magnetic fields.
Thus, a hint for the singularity
of the ground-state energy of the two-dimensional spin glass
may be obtained from the behavior of the singularities in the complex plane
as the width of strip is increased.

In \S 2, we explain the formulation of the present method
using examples of the width-two (ladder) systems.
Explicit formulas of the ground-state energy of the bond-random
and site-random models are given.
Investigation of systems with widths larger than two
needs additional techniques, which are explained in \S 3.
The exact ground-state energy is given for the case of width three.
We could obtain locations of singularities of the ground-state energy
only in limited regions of the complex probability plane
when the width is four and five.
This is due to limitations of computational resources (memory and CPU time).
We should stress here that exact analytical expressions,
not numerical values, of the ground-state energy
can be obtained if sufficient computational power is available.
In \S 4, behavior of singularities in the complex probability plane
is discussed.
Differences between bond-random and site-random models are described
and  their physical implications are discussed.

\section{Formulation of Exact Calculations}
{}~

In this section, we formulate the procedure for the evaluation
of the exact ground-state energy of the $\pm J$
and the site-random models.
The width-two (ladder) systems are treated as examples.
This formulation is based on the method proposed by
Derrida {\it et al.}~\cite{Derrida}
and Dress {\it et al.}~\cite{Dress}.
The main problem is how to evaluate averages over the
distribution of randomness.

\subsection{$\pm J$ model}
{}~

\label{sec21}
First, we consider the $\pm J$ Ising model on a ladder shown in
Fig.~\ref{ladderJ}.
The Hamiltonian is given by
\begin{equation}
\label{sgham}
{\cal H} = - \sum_{i=1}^{n} ( K_i \sigma_{i-1} \sigma_i
          +K'_i \sigma'_{i-1} \sigma'_i + L_i \sigma_i \sigma'_i )\:,
\end{equation}
where the variable $\sigma_i (\sigma'_i) = \pm 1 $ represents
an Ising spin on the upper (lower) line, and the exchange interactions
$K_i$, $K'_i$ and $L_i$ take the values $+1$ or $-1$ randomly with
the probabilities $p$ or $1-p$, respectively.
We calculate the ground-state energy per spin in the limit $n\to\infty$.

As is well-known, transfer-matrix calculation of the partition function of
Ising models is based on the constrained partition functions $Z_n^{\pm\pm}$,
in which the boundary spins are fixed as
$\sigma_n , \sigma'_n = \pm 1$ in the $2\times n$-spin ladder.
These constrained partition functions obey the following recursion relation
\begin{equation}
\label{recc}
\left[
\begin{array}{cc}
Z_{n+1}^{++}\\
Z_{n+1}^{+-}\\
Z_{n+1}^{-+}\\
Z_{n+1}^{--}
\end{array}
\right]
= T_{n+1}
\left[
\begin{array}{cc}
Z_n^{++}\\
Z_n^{+-}\\
Z_n^{-+}\\
Z_n^{--}
\end{array}
\right] ,
\end{equation}
using the transfer matrix $T_n$ given by
\begin{equation}
T_n =\left[
\begin{array}{cccc}
z^{ K_n+K'_n+L_n} & z^{ K_n-K'_n+L_n} & z^{-K_n+K'_n+L_n} & z^{-K_n-K'_n+L_n}\\
z^{ K_n-K'_n-L_n} & z^{ K_n+K'_n-L_n} & z^{-K_n-K'_n-L_n} & z^{-K_n+K'_n-L_n}\\
z^{-K_n+K'_n-L_n} & z^{-K_n-K'_n-L_n} & z^{ K_n+K'_n-L_n} & z^{ K_n-K'_n-L_n}\\
z^{-K_n-K'_n+L_n} & z^{-K_n+K'_n+L_n} & z^{ K_n-K'_n+L_n} & z^{ K_n+K'_n+L_n}
\end{array}
\right] ,
\end{equation}

\noindent where
\begin{equation}
\nonumber
z = e^{\beta} .
\end{equation}
In the zero-temperature limit $z\gg1$, we have only to consider
the leading term of $z$, namely,
\begin{equation}
\label{zerolim}
\left[
\begin{array}{cccc}
Z_n^{++}\\
Z_n^{+-}\\
Z_n^{-+}\\
Z_n^{--}
\end{array}
\right]
\approx
\left[
\begin{array}{llll}
A_n^{++} z^{x_n} \\
A_n^{+-} z^{x_n+2a_n}\\
A_n^{-+} z^{x_n+2b_n}\\
A_n^{--} z^{x_n+2c_n}
\end{array}
\right] .
\end{equation}
The exponent $x_n$ represents the energy of the strip of length $n$ with
the boundary condition $\sigma = \sigma' = + 1$.
The exponents $a_n$, $b_n$ and $c_n$ correspond to the differences
of energy between the systems with different fixed boundary conditions.

Since we are interested only in the energy,
we concentrate on the exponents $x_n$, $a_n$, $b_n$ and $c_n$.
For the asymptotic form~(\ref{zerolim}), the recursion relation~(\ref{recc})
is reduced to
\begin{subequations}
\begin{eqnarray}
\label{recn1}
x_{n+1} & = & K+K'+L+x_n+2M ,\\
\label{recn2}
a_{n+1} & = & \max(-K'-L,-L+a_n,-K-K'-L+b_n,\nonumber \\
        &   & -K-L+c_n)-M ,\\
\label{recn3}
b_{n+1} & = & \max(-K-L,-K-K'-L+a_n,\nonumber \\
        &   & -L+b_n,-K'-L+c_n)-M ,\\
\label{recn4}
c_{n+1} & = & \max(-K-K',-K+a_n,\nonumber \\
        &   & -K'+b_n,c_n)-M ,
\end{eqnarray}
\end{subequations}
with
\begin{equation}
\nonumber
M =\max(0,-K'+a_n,-K+b_n,-K-K'+c_n) \ ,
\end{equation}
and $K = K_{n+1}$, $K' = K'_{n+1}$ and $L = L_{n+1}$.
Because of the global reflection symmetry in the spin space,
$b_n$ is equal to $a_n$ and $c_n$ is equal to zero.
Then, the four recursion relations in
{}~(\ref{recn1})--(\ref{recn4}) are reduced to
the following two recursion relations
\begin{subequations}
\begin{eqnarray}
\label{recn5}
x_{n+1} & = & K+K'+L+x_n+2M' ,\\
\label{recn6}
a_{n+1} & = & \max(-K'-L,-L+a_n,-K-K'-L+a_n,\nonumber \\
        &   & -K-L)-M' ,
\end{eqnarray}
\end{subequations}
with
\begin{equation}
\nonumber
M' = \max(0,-K'+a_n,-K+a_n,-K-K') \ .
\end{equation}

The ground-state energy per spin is expressed as
\begin{eqnarray}
E(p) & = & \lim_{n\to\infty} \left. \frac{1}{2n}
           \left( -\frac{1}{\beta}\ln Z_n \right) \right|_{T=0} \nonumber\\
     & = &-\lim_{n\to\infty} \frac{1}{2n} \left[
           x_n + \max(0,2a_n,2a_n,0) \right]\nonumber\\
     & = &-\lim_{n\to\infty}\frac{1}{2n}\bigg[
           \sum_{m=0}^{n-1}(x_{m+1}-x_{m})
           +x_0 +\max(0,2a_n)\bigg]\nonumber\\
\label{Ep}
     & = & - \frac{1}{2} \ll x_{n+1} - x_n \gg \ ,
\end{eqnarray}
where the double brackets $\ll \cdots \gg$ represent the average
over the bond configuration.
The quantity $-(x_{n+1} - x_n)$ corresponds to the increase
of the energy by the transfer from $Z_n$ to $Z_{n+1}$.
Equation~(\ref{recn5}) means that the quantity $(x_{n+1} - x_n)$
is a function of $a_n$, $K_{n+1}$, $K'_{n+1}$ and $L_{n+1}$.
Since $K_{n+1}$, $K'_{n+1}$ and $L_{n+1}$ are independent of each other
and of $a_n$,
the probability $P(a_n,K_{n+1},K'_{n+1},L_{n+1})$ is decomposed as
$P(a_n)\cdot P(K_{n+1})\cdot P(K_{n+1})\cdot P(K_{n+1})$.
Among the four factors in this product, only the distribution $P(a_n)$
is nontrivial.
The function $P(a_n)$ can be calculated if we analyze the change of state
from the $n$th stage of transfer to the $(n+1)$th stage.
Details are explained in Appendix~\ref{apA}.
The result is
\begin{equation}
\label{exactEp}
E(p) = - \frac{3-3p+23p^2-40p^3+24p^4}{2(1+p+3p^2-8p^3+4p^4)} \ .
\end{equation}
This quantity is plotted as a function of real $p$ between $0$ and $1$
in Fig.~\ref{2xinf_bond}.
Note that this function is symmetric for the exchange of $p$ with $1-p$
as it should be.

\subsection{Site-random model}
{}~

\label{sec22}
Next, we consider the site-random Ising model on a ladder.
The Hamiltonian is given by the same form as in eq.~(\ref{sgham}),
except for the expressions of the exchange interactions
$K_n$, $K'_n$ and $L_n$.
The site-random model is composed of $A$ and $B$ ions with concentrations
$c$ and $1-c$, respectively.
The value of the exchange interaction is $+1$ for the neighboring ion pair
$A$--$A$ and is $-1$ for $A$--$B$ and $B$--$B$.

Calculations of the ground-state energy in this model
can be carried out quite similarly to those of the $\pm J$ model.
Formulas~(\ref{sgham}) to (\ref{Ep}) do not need modifications.
Only the evaluation of the final expression~(\ref{Ep}) requires
different treatments as explained below.

In the site-random model, the variables $K_{n+1}$, $K'_{n+1}$ and $L_{n+1}$
are functions of the type of ions $X_n$ and $X'_n$
which are either $A$ or $B$ in the $n$th row
and ions $X_{n+1}$ and $X'_{n+1}$ in the $(n+1)$th row.
Hence, the quantity $(x_{n+1}-x_n)$ is a function of
$a_n$, $X_n$, $X'_n$, $X_{n+1}$ and $X'_{n+1}$.
Let us remember that the variables $X_{n+1}$ and $X'_{n+1}$
located at the edge of the strip are random.
That is,
\begin{equation}
P(X) =
\left\{
\begin{array}{cc}
c & \mbox{ for } X = A \\
1-c & \mbox{ for } X = B
\end{array}
\right. ,
\end{equation}
where $X$ stands for $X_{n+1}$ or $X'_{n+1}$.
On the other hand, the variables $X_{n}$ and $X'_{n}$ are not random,
since they can be regarded as having already been fixed
in the previous step of transfer.
Then, the distribution function is written as
\begin{eqnarray}
\lefteqn{P(a_n,X_{n},X'_{n},X_{n+1},X'_{n+1})} \nonumber \\
& = & P(a_n,X_{n},X'_{n})\cdot P(X_{n+1})\cdot P(X'_{n+1}) \ .
\end{eqnarray}
The non-trivial part of distribution function is $P(a_n,X_n,X'_n)$.
Evaluation of this $P(a_n,X_n,X'_n)$ proceeds quite similarly as
in the case of the $\pm J$ model.
Details are found in Appendix~\ref{apA}.
We obtain the explicit expression of
the ground-state energy as
\begin{equation}
\label{exactEc}
E(c) = - \frac{3(1-3c^2+6c^3-3c^4)}{2(1+c-c^2)(1-c+c^2)} \ .
\end{equation}
The value of this energy is drawn as a function of real $c$
in Fig.~\ref{2xinf_site}.
It is seen that this quantity is symmetric under the exchange of
$c$ and $1-c$ as was discussed generally in Ref.~\cite{sristr}.

\section{Systems with Larger Widths}
{}~

In the present section, we generalize the method explained
in the previous section to strips with width lager than two for the $\pm J$ and
the site-random models.

In the case of width $3$,
the constrained partition functions in the low temperature limit is
expressed as
\begin{equation}
\left[
\begin{array}{cccc}
Z_n^{+++}\\
Z_n^{++-}\\
Z_n^{+-+}\\
Z_n^{+--}\\
Z_n^{-++}\\
Z_n^{-+-}\\
Z_n^{--+}\\
Z_n^{---}
\end{array}
\right]
\approx
\left[
\begin{array}{llll}
A_n^{+++} z^{x_n} \\
A_n^{++-} z^{x_n+2a_n}\\
A_n^{+-+} z^{x_n+2b_n}\\
A_n^{+--} z^{x_n+2c_n}\\
A_n^{-++} z^{x_n+2c_n}\\
A_n^{-+-} z^{x_n+2b_n}\\
A_n^{--+} z^{x_n+2a_n}\\
A_n^{---} z^{x_n}
\end{array}
\right] ,
\end{equation}
where the allocation of the exponents $x_n,a_n,b_n$ and $c_n$ takes
the global spin-flip symmetry into account.
These exponents satisfy the following recursion relations analogous to
eqs.~(\ref{recn5}),(\ref{recn6}):
\begin{subequations}
\begin{eqnarray}
x_{n+1} & = & K+K'+K''+L+L''+x_n+2M'' ,\\
a_{n+1} & = & \max(-K''-L',-L'+a,-K'-K''-L'+b,-K'-L'+c,\nonumber\\
        &   & -K-K''-L+c,-K-L'+b,-K-K'+a,\nonumber\\
        &   & -K-K'-K'')-M'' ,\\
b_{n+1} & = & \max(-K'-L-L',-K'-K''-L-L'+a,-L-L'+b,\nonumber\\
        &   & -K''-L-L'+c,-K-K'-L-L'+c,-K-K'-K''-L-L'+b,\nonumber\\
        &   & -K-L-L'+a,-K-K''-L-L'')-M'' ,\\
c_{n+1} & = & \max(-K'-K''-L,-K'-L+a,-K''-L+b,-L+c,\nonumber\\
        &   & -K-K'-K''-L+c,-K-K'-L+b,-K-K''-L+a,\nonumber\\
        &   & -K-L)-M'' ,
\end{eqnarray}
\end{subequations}

\noindent with
\begin{eqnarray}
M'' & = & \max(0,-K''+a,-K+b,-K'-K''+c,-K+c,-K-K''+b,\nonumber\\
    &   & -K-K'+a,-K-K'-K'') \ ,
\end{eqnarray}

\noindent where
$a = a_n$, $b = b_n$, $c = c_n$, $K = K_{n+1}$, $K' = K'_{n+1}$,
$K'' = K''_{n+1}$, $L = L_{n+1}$ and $L' = L'_{n+1}$.

We apply the above method to the $\pm J$ and the site-random models.
However, as was mentioned in Appendix~\ref{apA},
the same approach to the case of width $2$ requires
a symbolic inverse of a matrix of size $28\times28$ or $60\times60$,
which requires exceeding amount of CPU time.
The following trick makes it possible to calculate the exact ground-state
energy without explicit symbolic inverse of large matrices.

Equation~(\ref{dist}) can be written as
\begin{equation}
P_i \equiv P(a_n) = (\overline{S}^{-1})_{i,N} =
 \frac{\tilde{S}_{N,i}}{\det(\overline{S})} \ ,
\end{equation}
where  the index $i\ (= 1,\cdots,N)$ of $P_i$ stands for an element
of the set $\{a_n\}$
and $\tilde{S}_{N,i}$ denotes the co-factor of $\overline{S}$.
Then, eq.~(\ref{Ep}) is expressed explicitly as
\begin{eqnarray}
E(p) & = & - \frac{1}{2}\sum_{i} P_i \cdot (x_{n+1} - x_n)_i \nonumber\\
\label{avE}
& = & - \frac{1}{2 \det(\overline{S})}\sum_{i} \tilde{S}_{N,i} \cdot
 (x_{n+1} - x_n)_i \ .
\end{eqnarray}

Let us recall here that $\tilde{S}_{N,i}$ and $\det(\overline{S})$ are
polynomials of $p$ with integer coefficients.
Since it is easy to calculate $\tilde{S}_{N,i}$ or $\det(\overline{S})$ for
given $p$ numerically,
we can determine the coefficients of the polynomials
from numerical values at several points:
If the degree of the polynomial is $m$, numerical values of the polynomial at
$(m+1)$ points are sufficient to determine the exact values of coefficients.
The same technique holds for the site-random model.

In this way we have carried out the average of eq.~(\ref{Ep})
and obtained the ground-state energy as
\begin{equation}
\label{fracEp}
E(p) = - \frac{N(p)}{D(p)} \ ,
\end{equation}
with
\begin{eqnarray}
N(p) & = & 5-25p+158p^2-596p^3+1703p^4-3774p^5 \nonumber\\
     &   & +6322p^6-8368p^7+9892p^8-11536p^9 \nonumber\\
     &   & +12304p^{10}-10112p^{11}+5568p^{12} \nonumber\\
     &   & -1792p^{13}+256p^{14} ,\\
D(p) & = & 3(1-3p+22p^2-82p^3+231p^4 \nonumber\\
     &   & -516p^5+844p^6-960p^7 \nonumber\\
     &   & +720p^8-320p^9+64p^{10})
\end{eqnarray}
in the $\pm J$ model, and
\begin{equation}
\label{fracEc}
E(c) = - \frac{N_{SR}(c)}{D_{SR}(c)} \ ,
\end{equation}
with
\begin{eqnarray}
N_{SR}(c) & = & 5-15c^2+22c^3+19c^4-52c^5+40c^6 \nonumber\\
          &   & -112c^7+370c^8-580c^9+468c^{10} \nonumber\\
          &   & -192c^{11}+32c^{12} , \\
D_{SR}(c) & = & 3(1-c+3c^2-4c^3+6c^4-6c^5+2c^6) \nonumber\\
          &   & \times(1+c-3c^2+6c^4-6c^5+2c^6)
\end{eqnarray}
in the site-random model.
These results (\ref{fracEp}) and (\ref{fracEc}) are plotted
as functions of $p$ or $c$ in Figs.~\ref{3xinf_br} and \ref{3xinf_sr}.

In the case of width $4$,
the number of the bases of the transition matrix $S$
defined in Appendix~\ref{apA}
exceeds two hundred.
Accordingly the degrees of polynomials of the denominator and numerator
of the expression of the energy (\ref{avE}) are much larger
than in the case of width $3$.
Thus we have to determine the coefficients of polynomials
of very large degrees from numerical values of the polynomials
at many points.
This procedure requires prohibitively high-precision calculations,
which we were not able to perform.

Instead, we determined the locations of singularities
(where $\det(\overline{S}) = 0$)
in restricted regions of the complex probability plane.
First,
the value of $\det(\overline{S})$
is scanned in the complex plane with the interval of $0.01$
for real and imaginary parts of $p$ or $c$.
Zeros of $\det(\overline{S})$ correspond to the crossing points
of two sets of curves,
the first one corresponding to $\mbox{Re}[\det(\overline{S})] = 0$
and the second to $\mbox{Im}[\det(\overline{S})] = 0$.
These curves are displayed in Fig.~\ref{curve}.
Thus, the regions in which zeros are likely to be located are specified
from the crossing points.
Next,
the zeros in the specified regions are precisely evaluated numerically using
the subroutine of finding zeros of a function in the complex plane.
In some cases, zeros of $\det(\overline{S})$ coincide with zeros
of the numerator of eq.~(\ref{avE}).
In such cases the energy is not singular.
This type of zeros of $\det(\overline{S})$ should be excluded from
our considerations.
Direct numerical evaluation of the energy at such points reveals
whether or not a zero of $\det(\overline{S})$ is a real singularity
of the energy.
We have in this way singled out real singularities of the energy.
Note that the present method gives much more precise results than
directly locating divergences of the energy in the complex plane
which accompanies large rounding errors.

In the present analyses, we have concentrated on the region
$-0.5\leq\mbox{Re}(p,c)\leq1.5$ and $-1\leq\mbox{Im}(p,c)\leq1$
due to the limit of computational time.
Although some singularities lie outside this region,
physically important ones are located near the real axis in the range
$0\leq p,c \leq1$.
Therefore, for the purpose of investigation of effects of singularities
in real physical systems,
it is sufficient to concentrate our attention to the above mentioned region
in the complex plane.

In the case of width 5,
we could locate only one singularity nearest to the range $0\leq p\leq1$
for the $\pm J$ model by using the same method as the case of width 4.
The site-random model was not investigated for the width 5.

Let us note that the unit of strips is not limited to squares.
For example, a strip composed of triangles can be treated,
as shown explicitly for the case of width 2 in Appendix~\ref{apB}.

\section{Summary and Discussion}
{}~

In the present paper, we have generalized the method for the calculation
of the exact ground-state energy proposed
by Derrida {\it et al.}~\cite{Derrida} and Dress {\it et al.}~\cite{Dress}.
The ground-state energy of finite-width strips
can be obtained by using this method.

We have applied the method to the $\pm J$ Ising model
on strips with width $2$ to $5$ and the site-random Ising model
with width $2$ to $4$.
In the case of widths $2$ and $3$, the exact ground-state energy
and the singularities in the complex-probability plane
of randomness have been obtained explicitly.
In the case of width $4$ and $5$, the exact location of singularities
near the physical region ($0\leq p,c\leq1$) have been identified
by numerical calculation.

In the $\pm J$ model, Fig.~\ref{singular}(a), there seems to exist a series
of singularities
which may approach a limiting point around $p \sim 0.9$ on the real axis
as the size of the strip grows.
In the site-random model, Fig.~\ref{singular}(b),
such series does not seem to exist within the present calculations.
These results suggest that the ground-state energy of the $\pm J$ model may
have a singularity at the critical concentration
$p_{\rm c} \sim 0.89$,~\cite{2DIs}
and that the ground-state energy of the site-random model does not have any
singularity at the critical concentration
$c_{\rm c} \sim 0.63$~\cite{sristr}.
Further investigations are necessary to clarify the significance of these
observations.

\section*{Acknowledgements}
{}~

We would like to thank Dr. Y.~Ozeki and Professor S.~Kobe
for helpful comments.
One of the present authors (Y.~N.) is grateful for the financial support
of the Japan Society for the Promotion of Science
for Japanese Junior Scientists.

\appendix
\section{Evaluation of the Functions $P(a_n)$ and $P(a_n,X_n,X'_n)$}
\label{apA}
{}~

The recursion relation~(\ref{recn6}) has a stationary distribution $P(a_n)$.
As has been mentioned in \S~{\ref{sec21}},
the explicit form of this distribution function is necessary for exact
calculation of the ground-state energy.
It is found easily by explicit iteration
of the recursion relation~(\ref{recn6}) starting from various finite values
of $a_n$ that the set $\{a_n\}$ is restricted to a closed finite one.
That is, we find $\{a_n\} = \{ 2, 1, 0, -1, -2 \}$
in the $\pm J$ model of width $2$.
Similarly, in the site-random model with width $2$,
it is  necessary to obtain the joint distribution function
$P(a_n,X_n,X'_n)$ as explained in \S~{\ref{sec22}}.
It is again found by explicit iteration
that the set $\{a_n,X_n,X'_n\}$ consists of $(0,A,A)$, $(-1,A,A)$,
$(-2,A,A)$, $(0,A,B)$, $(-1,A,B)$, $(-2,A,B)$, $(0,B,A)$, $(-1,B,A)$,
$(-2,B,A)$, $(2,B,B)$, $(1,B,B)$ and $(0,B,B)$
as the stationary state of the recursion relation~(\ref{recn6}).

Bearing these observations in mind, we construct a matrix $S$
which transforms $a_n$ to $a_{n+1}$.
In the case of the $\pm J$ model with width $2$, we find from eq.~(\ref{recn6})
\begin{eqnarray}
|a_{n+1}=+2\rangle \:\:\: |a_{n+1}=+1\rangle \:\:\:
|a_{n+1}= 0\rangle \:\:\: |a_{n+1}=-1\rangle \:\:\: |a_{n+1}=-2\rangle
\qquad \qquad \qquad && \nonumber\\*
S=\left[\begin{array}{ccccc}
p^{2}q+q^{3} & p^{2}q+q^{3} & 0 &
2pq^{2} & 2pq^{2} \\
0 & 0 & p^{2}q+2pq^{2}+q^{3} & 0 & 0 \\
p^{3}+3pq^{2} & p^{3}+3pq^{2} & 0 &
3p^{2}q+q^{3} & 3p^{2}q+q^{3} \\
0 & 0 & p^{3}+2p^{2}q+pq^{2} & 0 & 0 \\
2p^{2}q & 2p^{2}q & 0 &
p^{3}+pq^{2} & p^{3}+pq^{2} \\
\end{array}\right]\begin{array}{l}
|a_{n}=+2\rangle \\
|a_{n}=+1\rangle \\
|a_{n}= 0\rangle \ , \\
|a_{n}=-1\rangle \\
|a_{n}=-2\rangle
\end{array}
\label{transit}
\end{eqnarray}

\noindent with $q = 1-p$.
The stationary distribution $P(a_n)$ is given by the eigenvector of $S$
with the eigenvalue unity:
\begin{equation}
\label{a2}
\left(
\begin{array}{c}
P(2)\\
\vdots\\
P(-2)
\end{array}
\right)
= S
\left(
\begin{array}{c}
P(2)\\
\vdots\\
P(-2)
\end{array}
\right) ,
\end{equation}
where $P(a_n)$ satisfies
\begin{equation}
\label{a3}
\sum_{\{a_n\}} P(a_n) = 1 \ .
\end{equation}
It is convenient to introduce the matrix $\overline{S}$ defined by
\begin{equation}
\overline{S} = \left[
\begin{array}{ccccc}
1-S_{1,1}   &  -S_{1,2}   & \cdots &  -S_{1,N-1}   &  -S_{1,N}   \\
 -S_{2,1}   & 1-S_{2,2}   & \cdots &  -S_{2,N-1}   &  -S_{2,N}   \\
\vdots      & \vdots      & \ddots & \vdots        & \vdots      \\
 -S_{1,N-1} &  -S_{2,N-1} & \cdots & 1-S_{N-1,N-1} &  -S_{N-1,N} \\
1           & 1           & \cdots & 1             & 1
\end{array}
\right] ,
\end{equation}

\noindent where $N$ denotes the size of the set $\{a_n\}$
($N = 5$ in the case of eq.~(\ref{transit})).
Then eqs.~(\ref{a2}) and (\ref{a3}) are written in a single relation
\begin{equation}
\overline{S}
\left(
\begin{array}{c}
P(2)\\
\vdots\\
\vdots\\
P(-2)
\end{array}
\right)
=
\left(
\begin{array}{c}
0\\
\vdots\\
0\\
1
\end{array}
\right) .
\end{equation}
Thus, we obtain
\begin{equation}
\left(
\begin{array}{c}
P(2)\\
\vdots\\
\vdots\\
P(-2)
\end{array}
\right)
= \overline{S}^{-1}
\left(
\begin{array}{c}
0\\
\vdots\\
0\\
1
\end{array}
\right)
=
\left(
\begin{array}{c}
(\overline{S}^{-1})_{1,N}\\
\vdots\\
\vdots\\
(\overline{S}^{-1})_{N,N}
\end{array}
\right) .
\label{dist}
\end{equation}

Similarly, in the case of the site-random model with width 2,
the distribution function can be obtained by using
the transition matrix
\begin{equation}
S=\left[\begin{array}{cccccccccccc}
 0 & 0 & 0 & 0 & 0 & 0 & 0 & 0 & 0 & 0 & {q^2} & {q^2} \\
 {q^2} & 0 & 0 & {q^2} & 0 & 0 & {q^2} & 0 & 0 & {q^2} & 0 & 0 \\
 0 & {q^2} & {q^2} & 0 & {q^2} & {q^2} & 0 & {q^2} & {q^2} & 0 & 0 & 0 \\
 0 & 0 & 0 & 0 & pq & pq & 0 & 0 & 0 & 0 & 0 & 0 \\
 pq & 0 & 0 & pq & 0 & 0 & pq & 0 & 0 & pq & 0 & 0 \\
 0 & pq & pq & 0 & 0 & 0 & 0 & pq & pq & 0 & pq & pq \\
 0 & 0 & 0 & 0 & 0 & 0 & 0 & pq & pq & 0 & 0 & 0 \\
 pq & 0 & 0 & pq & 0 & 0 & pq & 0 & 0 & pq & 0 & 0 \\
 0 & pq & pq & 0 & pq & pq & 0 & 0 & 0 & 0 & pq & pq \\
 0 & {p^2} & {p^2} & 0 & 0 & 0 & 0 & 0 & 0 & 0 & 0 & 0 \\
 {p^2} & 0 & 0 & {p^2} & 0 & 0 & {p^2} & 0 & 0 & {p^2} & 0 & 0 \\
 0 & 0 & 0 & 0 & {p^2} & {p^2} & 0 & {p^2} & {p^2} & 0 & {p^2} & {p^2}
\end{array}\right] .
\label{transit2}
\end{equation}

The matrix $S$ or $\overline{S}$ depends on models.
The size of the matrix increases exponentially as the system size increases.
The sizes of the matrices are $5$, $28$, $286$, $3400$
for widths $2$, $3$, $4$, $5$
in the $\pm J$ model, and $12$, $60$, $528$
for widths $2$, $3$, $4$ in the site-random model.
Thus the explicit evaluation of an inverse matrix by symbolic manipulation
becomes difficult.

\section{Strip Consisting of Triangles}
\label{apB}
{}~

We have calculated the ground-state energy of the one-dimensional $\pm J$
model on the linear triangular chain (Fig.~\ref{triangle})
by using the same method.
The ground-state energy is given by
\begin{equation}
E(p) = - \frac{2(2-2p+6p^2-4p^3)}{1+3p^2+2p^3} \ .
\end{equation}

This model has already been ``solved'' by
Fechner and B{\l}aszyk~\cite{Fechner}.
Although this expression is consistent with their Fig.~3,
their expression of the ground-state energy is neither consistent with
this expression nor with their own figure.

\newpage
\section*{Figure Captions}
{}~

\begin{figure}[h]
\vspace{-5ex}
\caption{
The ladder of the $\pm J$ model with random ferromagnetic (solid line)
and antiferromagnetic (dashed line) interactions.
}
\label{ladderJ}
\caption{
The ground-state energy of the ladder system of the $\pm J$ model
as a function of real $p$.
}
\label{2xinf_bond}
\caption{
The ground-state energy of the ladder system of the site-random model
as a function of real $c$.
}
\label{2xinf_site}
\caption{
The ground-state energy of the $\pm J$ model
on the $3\times\infty$-spin strip.
}
\label{3xinf_br}
\caption{
The ground-state energy of the site-random model
on the $3\times\infty$-spin strip.
}
\label{3xinf_sr}
\caption{
Curves on which the real part of $\det(\overline{S})$ vanishes,
(a) and (c), and the imaginary part vanishes, (b) and (d),
for the  $4\times\infty$-spin strips.
The figures (a) and (b) are for the $\pm J$ model,
and (c) and (d) are for the site-random model.
Only the quarter of the whole region
($\mbox{Re}(p,c)\geq0.5$, $\mbox{Im}(p,c)\geq0$)
is shown, because the models are symmetric under the exchange
$p,c \leftrightarrow p^{\ast},c^{\ast}$ (complex conjugate)
or $p,c \leftrightarrow 1-p,1-c$.
}
\label{curve}
\caption{
Singularities of the ground-state energy
in the $\pm J$ model (a) for the $2\times\infty$- to $5\times\infty$-spin
strips,
and in the site-random model (b) for the $2\times\infty$- to
$4\times\infty$-spin strips.
The symbols {\protect\large$\circ$}, $\Box$, $\Diamond$ and $\triangle$
correspond to the singularities of the
$2\times\infty$-, $3\times\infty$-, $4\times\infty$-, and $5\times\infty$-spin
strips, respectively.
The symbol {\protect\large$\bullet$}
in (a) denotes the transition point from the ferromagnetic phase to
a non-ferromagnetic phase along the real axis for
the $\pm J$ model, $p_{\rm c} = 0.89 \pm 0.01$~{\protect\cite{2DIs}}.
Only the quarter of the whole region is shown because of
the symmetry as explained in relation to Fig.~{\protect\ref{curve}}.
}
\label{singular}
\caption{
The linear triangular chain with random ferromagnetic (solid line) and
antiferromagnetic (dashed line) interactions.
}
\label{triangle}
\end{figure}

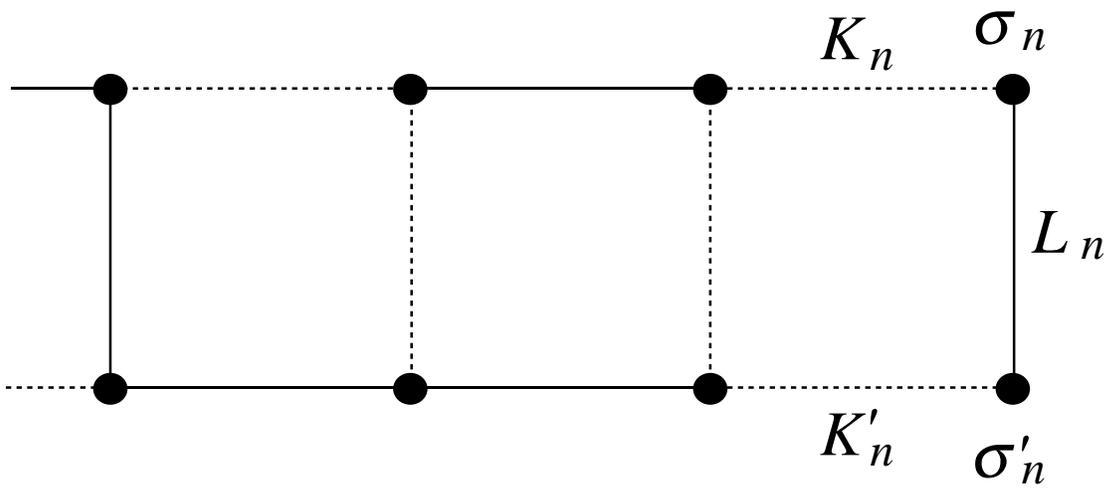
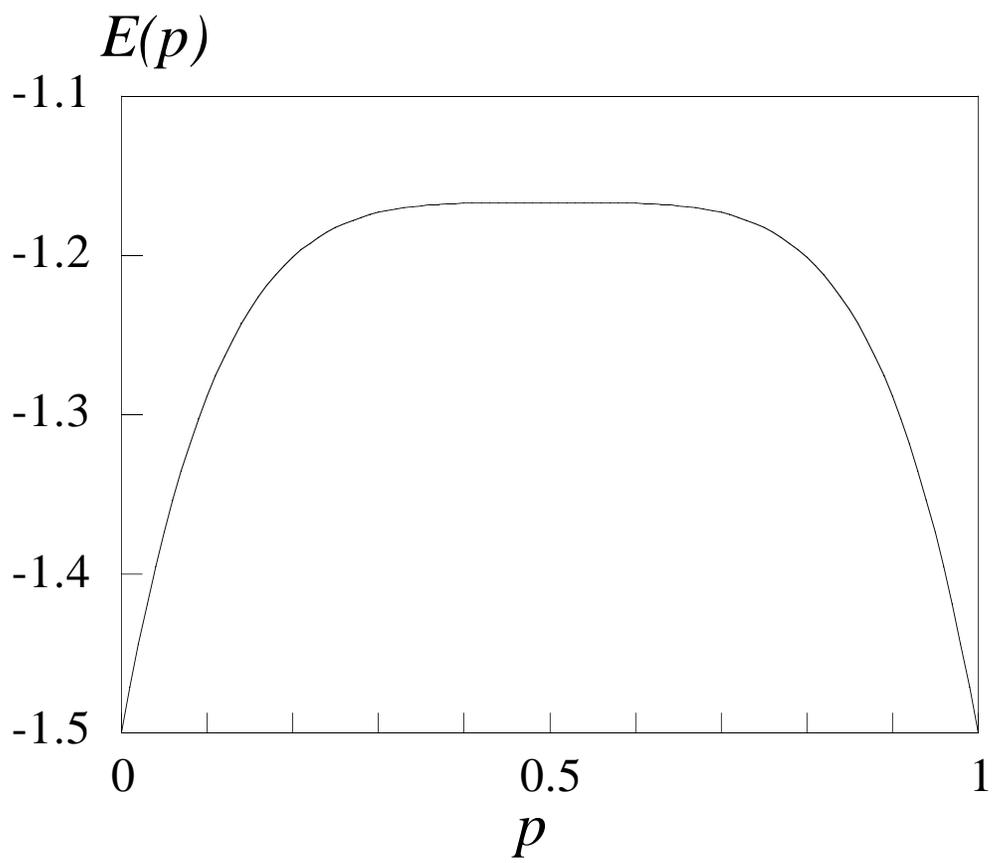
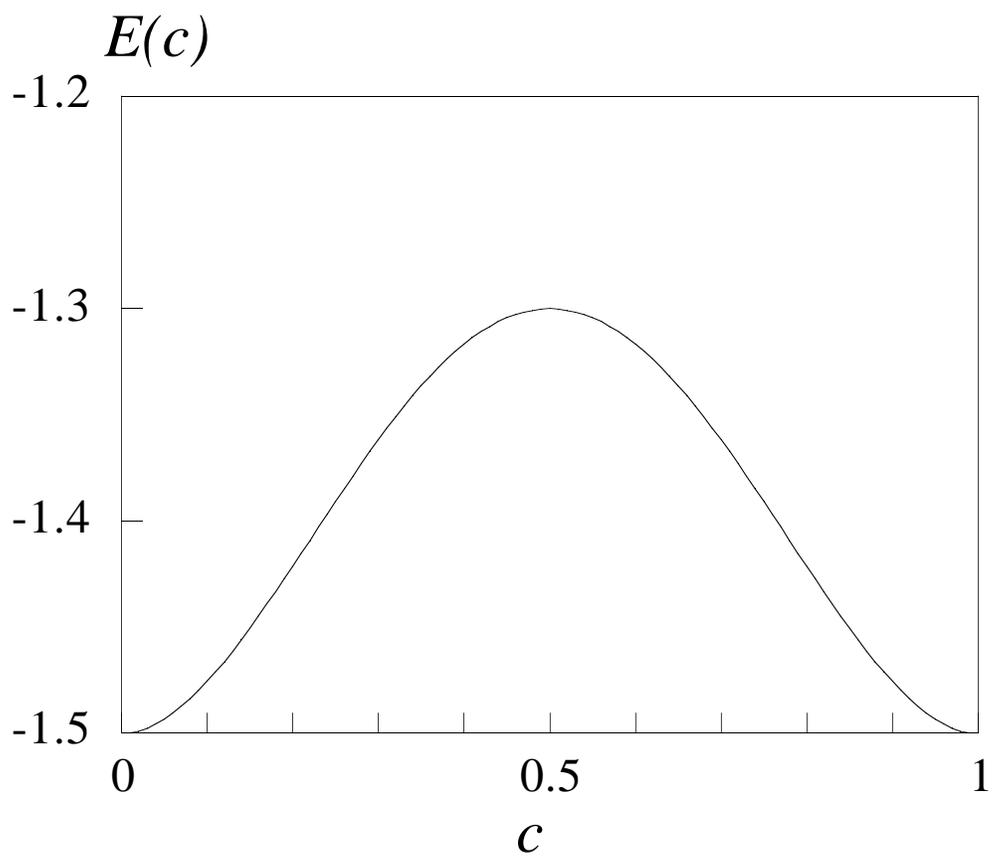
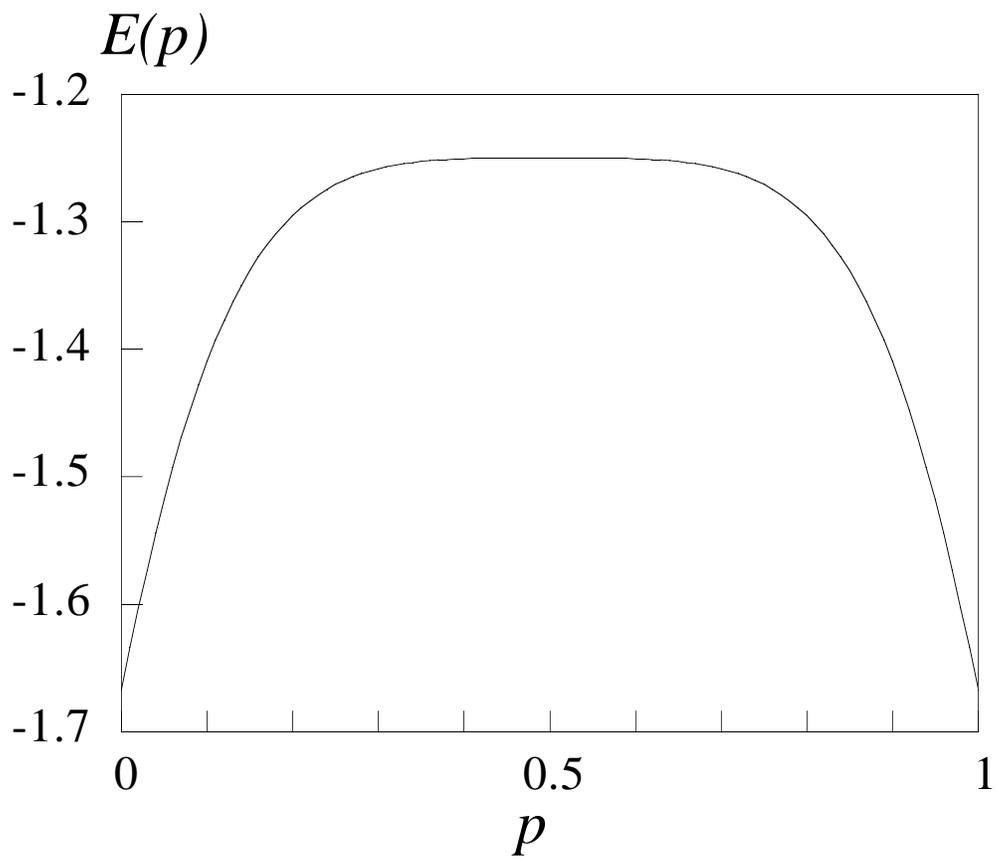
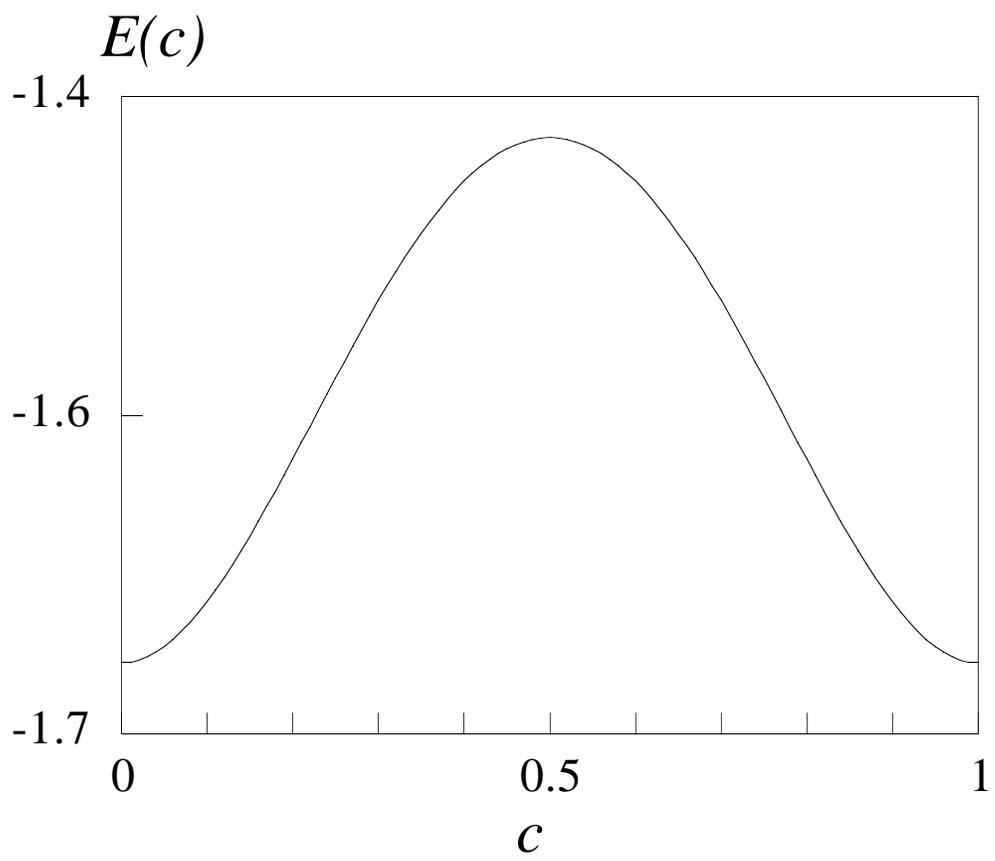
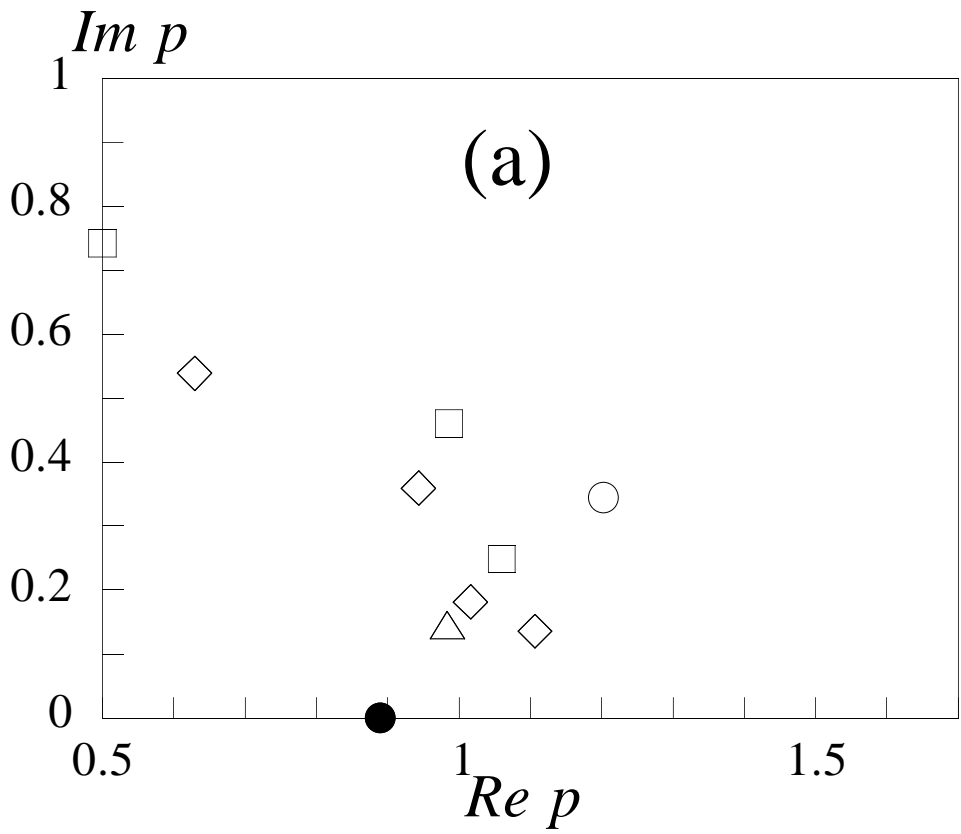
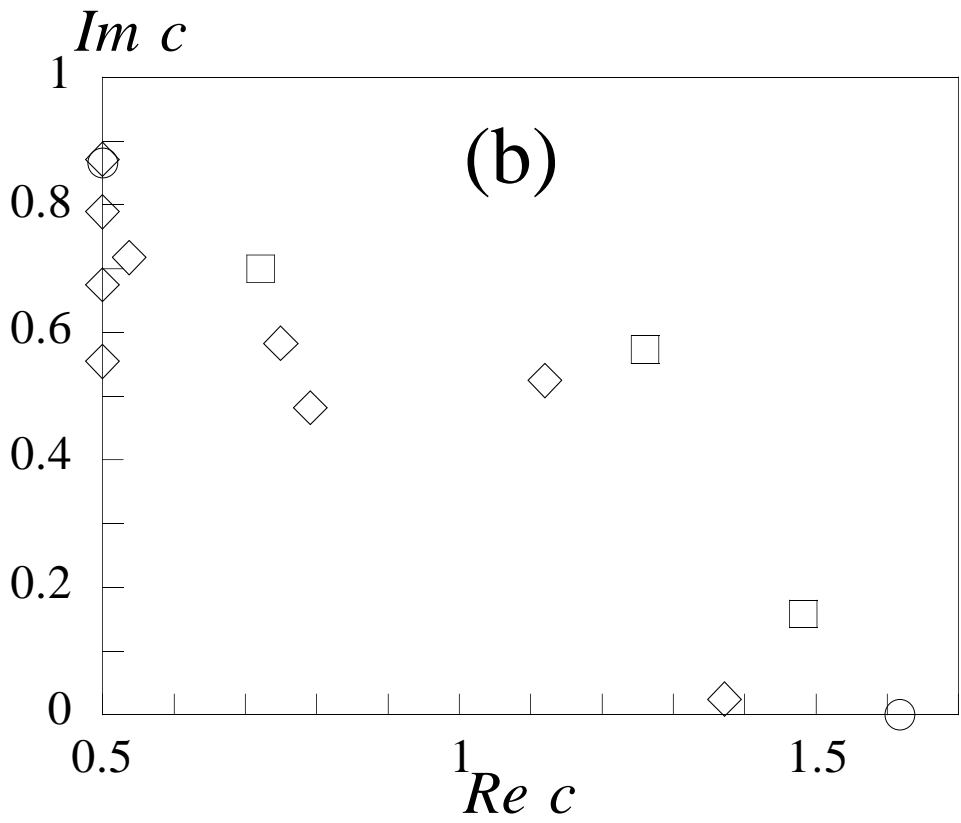
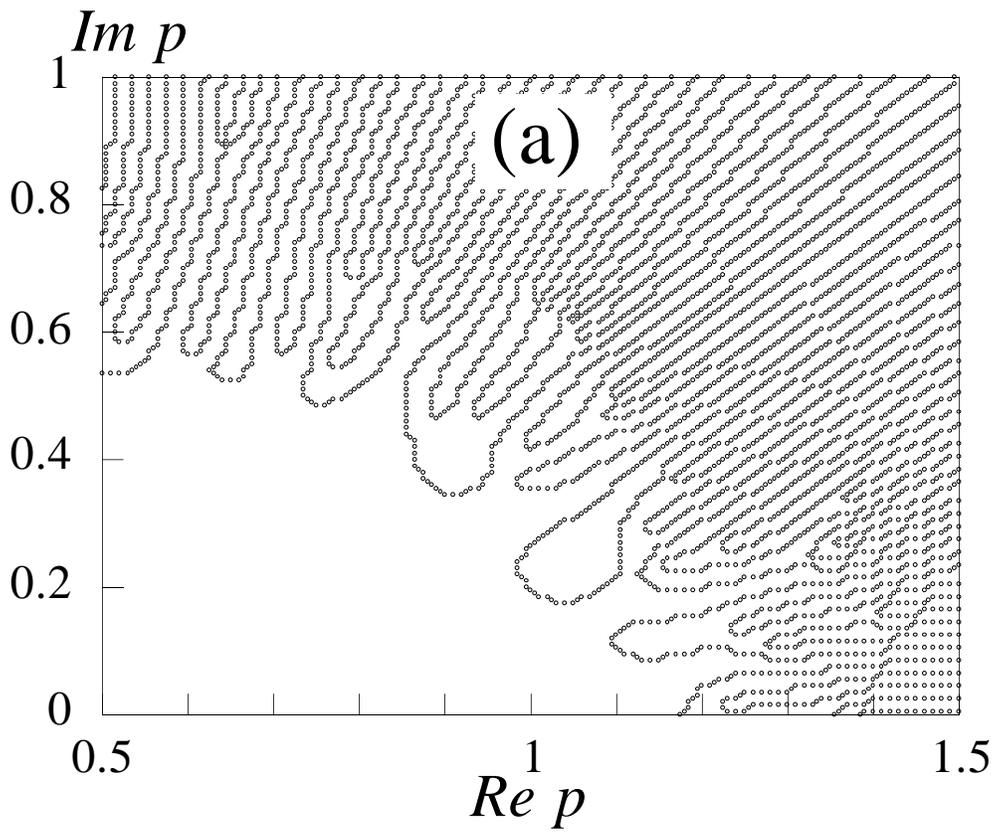
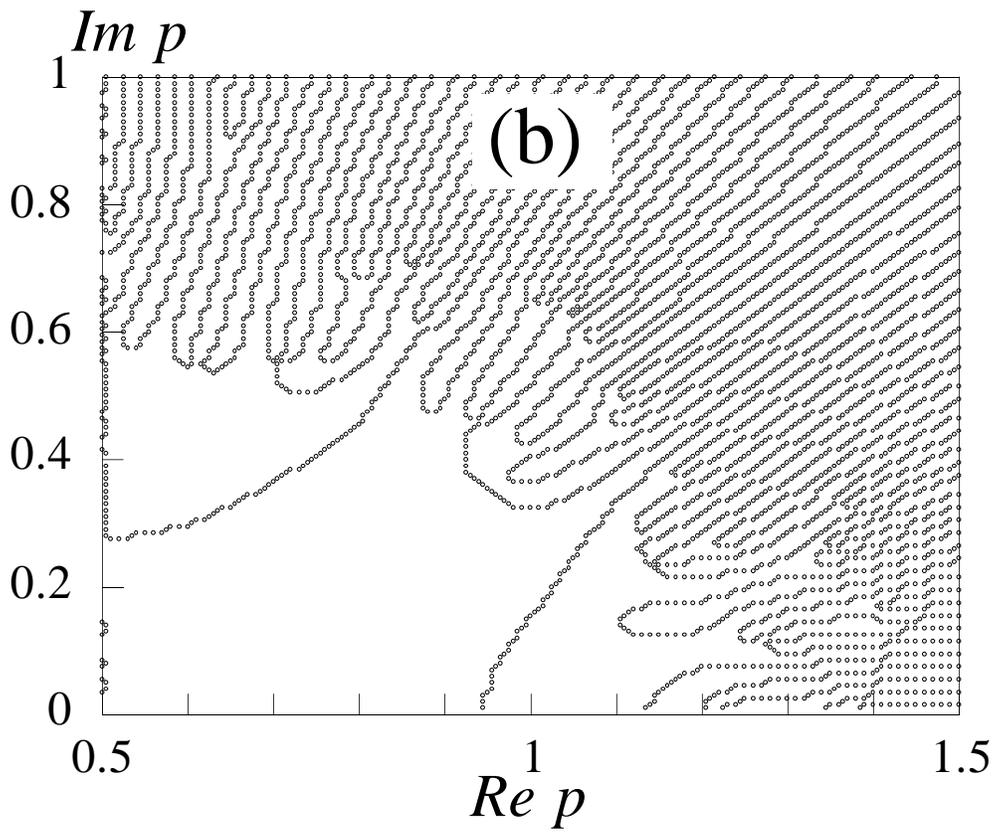
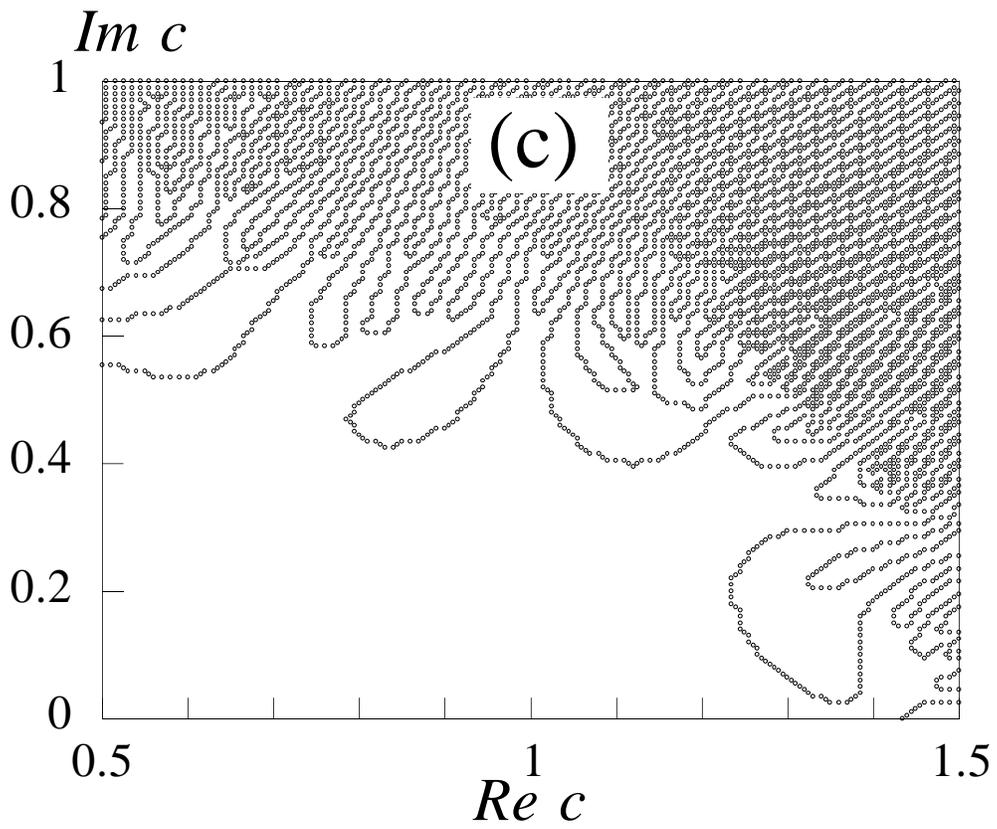
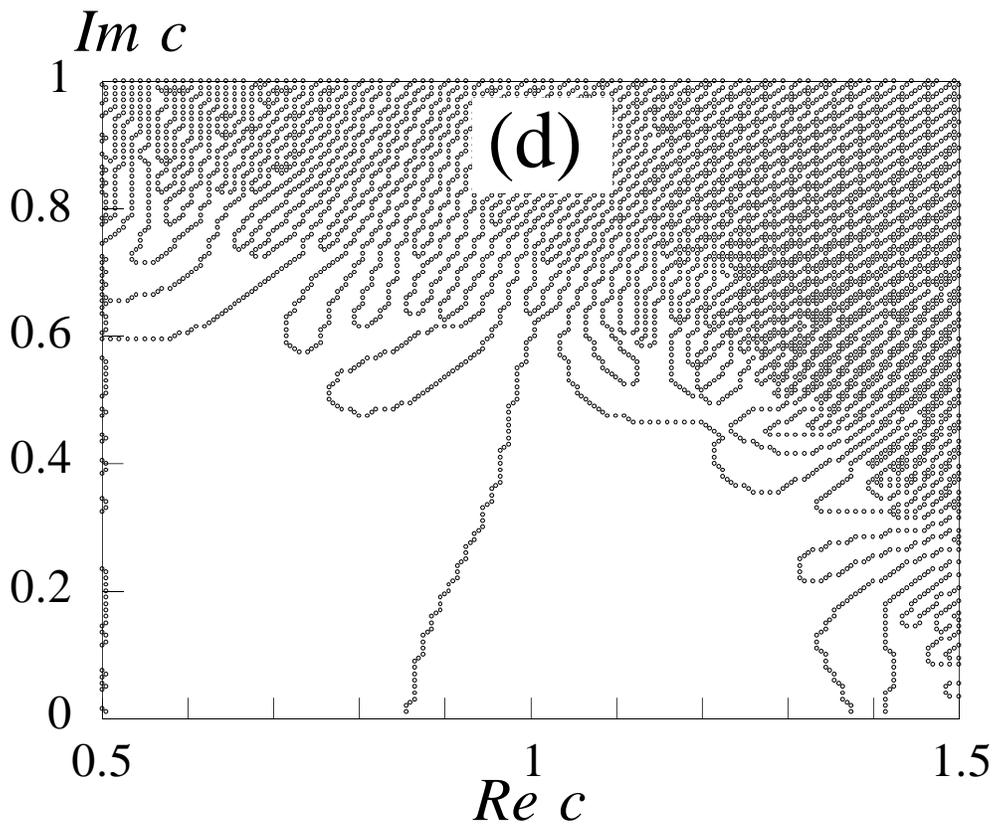
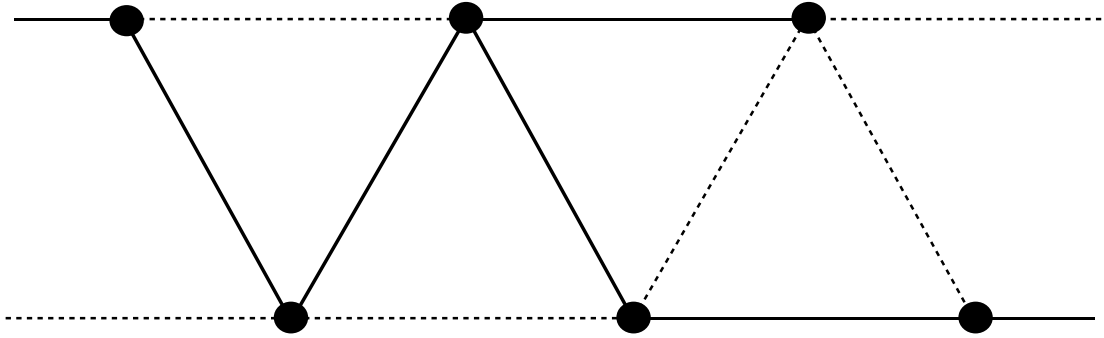
\end{document}